\documentstyle[preprint,aps,prl]{revtex}
\begin{document}
\preprint
\widetext
\title{Magnetic phase diagram of the Hubbard  model}
\author{
J. K. Freericks$^{(a),\dagger}$ and Mark Jarrell$^{(b)}$.
}
\address{
$^{(a)}$Department of Physics,
University of California, Davis, California 95616\\
$^{(b)}$Institute for Theoretical Physics,
University of California, Santa Barbara CA,\\
 and,\\
Department of Physics,
University of Cincinnati, Cincinnati, OH 45221
}
\date{\today}
\maketitle
\widetext
\begin{abstract}
The competition between commensurate and incommensurate
spin-density-wave phases in the infinite-dimensional single-band Hubbard
model is examined with quantum Monte Carlo simulation and
strong and weak coupling approximations.  Quantum fluctuations
modify the weak-coupling phase diagram by factors of order
unity and produce remarkable agreement with the quantum
Monte Carlo data, but strong-coupling theories (that map onto
effective Falicov-Kimball models) display pathological behavior.
The single-band model can be used to describe much of the 
experimental data in Cr and its dilute alloys with V and Mn.
\end{abstract}
\pacs{Pacs: }
\narrowtext
\paragraph*{Introduction.}
Spin-density-wave (SDW) order, in which the modulation wavevector
of the SDW is incommensurate with the underlying lattice, is one
of the most fascinating ordered states found in nature. 
Incommensurate magnetism occurs in both metallic and insulating
phases and on both frustrated and unfrustrated lattices.
In general, incommensurate magnetic order may be driven by either
frustration or by Fermi surface nesting 
with a wavevector that lies away from commensurate wavevectors.  It is 
important to understand which process plays a more important role, and 
to understand how many-body effects modify the stability of incommensurate
phases.  Here the effect of nesting is examined on an unfrustrated lattice
with strongly correlated electrons.
The resulting phase diagram is then compared to approximate results in the 
weak- and strong-coupling limits (Figs.~2 and 3).  Finally, our theoretical
results are compared to those found in Cr and its dilute alloys.

Elemental Cr is a paradigm for an antiferromagnetic (AF) metal~\cite{fawcett}
with incommensurate order driven by Fermi-surface nesting.  The lattice 
structure of Cr is an unfrustrated body-centered-cubic structure which may 
be modeled by a Hubbard model~\cite{hubbard} near half 
filling with moderate electron-electron correlations.
Adding electrons to Cr (by alloying with Mn) rapidly makes the 
magnetic order commensurate with the lattice, whereas removing electrons
from the system (by alloying with V) rapidly increases the incommensuration
and decreases the magnetic transition temperature, eventually to 
zero~\cite{mackintosh}.  

Heretofore, incommensurate magnetic order has mainly been examined within
the Hartree-Fock (HF) (weak-coupling) approximation that neglects
quantum fluctuations.  Penn~\cite{penn} found incommensurate order
in the three-dimensional Hubbard model, and Schulz~\cite{schulz}
found evidence for incommensurate phases on a square lattice.  

\paragraph*{Formalism.}
In this contribution the magnetic phase diagram of the single-band
Hubbard model is investigated in the limit of infinite
dimensions~\cite{metzner_vollhardt}.  This limit is useful, because
it has been shown to contain most of the physics expected of
three-dimensional Hubbard models, and the many-body effects can be
treated essentially exactly 
with the quantum Monte Carlo (QMC) techniques of Hirsch
and Fye~\cite{hirsch_fye}. This allows us to
demonstrate the existence of incommensurate order at finite
temperatures in a model that only includes Fermi-surface
nesting effects and electron-electron correlations.  We find
incommensurate magnetism to be a ubiquitous property of
this model.

The Hubbard model~\cite{hubbard} is described by the following 
Hamiltonian:
\begin{equation}
H=-{t^*\over 2\sqrt{d}}\sum_{\langle ij\rangle
\sigma}[c_{i\sigma}^{\dagger}c_{j\sigma}+
c_{j\sigma}^{\dagger}c_{i\sigma}]+U\sum_i
(n_{i\uparrow}-{1\over 2})(n_{i\downarrow}-{1\over 2})-
\mu\sum_{i\sigma}n_{i\sigma}\quad ,
\label{eq: hubbham}
\end{equation}
where $c_i^{\dagger}$ ($c_i$) is a creation (destruction) operator
for an electron at site $i$ with spin $\sigma$.  The hopping
matrix elements connect nearest neighbors on a hypercubic lattice
in $d$-dimensions; its magnitude is written as $t=t^*/2\sqrt{d}$
[to have a well-defined limit in infinite dimensions ($d\rightarrow
\infty$)].  All energies are expressed in units of the rescaled
hopping matrix element $t^*$.  The Coulomb repulsion is represented
by $U$ and the chemical potential by $\mu$.

In the limit of infinite dimensions the local approximation becomes
exact.  The electronic Green's function $G(i\omega_n)\equiv G_n$
is represented by an integral over the noninteracting density of states 
$\rho(y)=\exp(-y^2)/\sqrt{\pi}$
\begin{equation}
G_n=\int_{-\infty}^{\infty} dy {\rho
(y)\over{i\omega_n-\mu-\Sigma_n-y}}\equiv
F_{\infty}(i\omega_n+\mu-\Sigma_n)\quad\,,
\label{eq: gdef}
\end{equation}
with $\Sigma_n\equiv\Sigma(i\omega_n)$
the electronic self energy.  The magnetic susceptibility satisfies
Dyson's equation
\begin{equation}
\chi_{mn}({\bf q})=\chi_m^0({\bf q})\delta_{mn}-T\sum_r
\chi_m^0({\bf q})\Gamma_{mr}\chi_{rn}({\bf q})\quad ,
\label{eq: chidef}
\end{equation}
with $\Gamma_{mr}\equiv\Gamma(i\omega_m,i\omega_r)$ the (local)
irreducible vertex function in the SDW channel.  The bare
susceptibility $\chi_m^0({\bf q})$ is defined by
\begin{eqnarray}
\chi_m^0({\bf q})&\equiv& -T\sum_k G_m({\bf k})G_m({\bf k}+{\bf q})\cr
&=&-{1\over\sqrt{\pi}}{1\over\sqrt{1-X^2}}\int_{-\infty}^{\infty}
dy {e^{-y^2}\over i\omega_n-\mu-\Sigma_n-y}F_{\infty}\left [ 
{{i\omega_n+\mu-\Sigma_n-Xy\over\sqrt{1-X^2}}} \right ] \quad.
\label{eq: chi0def}
\end{eqnarray}
The bare susceptibility only depends on the scalar parameter
$X({\bf{q}})\equiv\sum_{i=1}^d\cos {\bf q}_i/d$, which defines an equivalence
class of wavevectors in the infinite-dimensional Brillouin
zone~\cite{mueller_hartmann}.  $X({\bf{q}})$ can be parametrized
by the line that extends along the Brillouin zone diagonal
from the zone center ($X=1$) to the zone corner $(X=-1)$.  The
self energy and irreducible vertex function are extracted from
the self-consistent QMC simulations as described
previously~\cite{jarrell} (for each value of $U$, we employ a constant 
small value of $\Delta\tau$ for the spacing of the imaginary time slices 
in the Trotter breakup).

It is important to qualify the region of validity of different
approximation techniques by comparing them with the essentially
exact QMC results.  Here we compare both weak-coupling and
strong-coupling perturbation theories.  

In the weak coupling limit a renormalized Hartree-Fock
approach\cite{vandongen} is employed.  The N\'eel temperature is 
determined by the Stoner criterion
\begin{equation}
{1\over U}=\chi^0(X,T)\equiv T\sum_n\chi_n^0(X)\quad ,
\label{eq: stoner}
\end{equation}
where the bare particle-hole susceptibility
$\chi_n^0(X)$ is calculated with noninteracting Green's
functions $G_n^0$ [$\Sigma_n=0$ in Eq.~(\ref{eq: gdef})].
This HF transition temperature is reduced by factors of order three
due to quantum fluctuations~\cite{vandongen}, even in the limit
$U/t^*\rightarrow 0$.  Quantum fluctuations modify the Stoner
criterion to~\cite{freericks_weak}
\begin{equation}
{1\over U}=\chi^0(X,T)-\chi^0(X=0,T)=\chi^0(X,T)+T\sum_n (G_n^0)^2
\quad ,
\label{eq: mod_stoner}
\end{equation}
in the limit $U/t^*\rightarrow 0$.  These fluctuations initially
reduce $T_c$ by the factor $\exp [-\chi^0(0,T_c)/\rho (\mu)]$ in the
weak-coupling limit~\cite{vandongen,freericks_weak}.

In the strong-coupling limit the Hubbard model can be mapped onto
a Falicov-Kimball model~\cite{janis,li}.  This mapping is exact
in the limit $U/t^*\to\infty$ (which reproduces the atomic limit)
for the self energy  but is not exact for the irreducible vertex functions 
in this limit.  As a result, the strong-coupling theories display pathological 
behaviors.  More explicitly, these approximations assume that the 
down-spin particles form a {\it static} background when the up-spin 
particles move and {\it vice versa}; this system is then
described by Falicov-Kimball models\cite{falicov_kimball} for both the
spin-up and spin-down electrons that are self-consistently coupled
together.  Two different self-consistent coupling schemes have been
proposed so far~\cite{janis,li}.  Jani\v s and Vollhardt's approximation~\cite{janis} underestimates
the SDW susceptibility at half filling which strongly
suppresses $T_c$ and does not reproduce the
Heisenberg limit of $T_c\approx t^{*2}/2U$.  Li and d'Ambrumenil's
approximation~\cite{li} is correct for large $U$ at half filling, but has 
the pathological behavior of predicting ferromagnetism away from
half filling (with a finite transition temperature in the
$U/t^*\rightarrow\infty$ limit) because of segregation in the
effective Falicov-Kimball model.  This latter pathology occurs because the
zero temperature occupation number of the static particles is
$0$, $0.5$, or $1$, and segregation occurs in the Falicov-Kimball model
whenever the static particle concentration is $0.5$, and does not
equal the mobile particle concentration~\cite{freericks_fk}.  We deal 
with this pathology, by only considering ordered states with $X<0$.

\paragraph*{Results.}
To determine the magnetic transition temperatures, we calculated the
magnetic susceptibility for all $X({\bf{ q}})$ in the Brillouin zone.
As shown in Fig.~1, the susceptibility always displayed a maximum
at a distinct value of $X$.  The transition temperature was then
inferred from interpolation (or extrapolation) of the peak 
inverse susceptibility, as shown in the inset to Fig.~1. 
At half filling~\cite{jarrell} the Hubbard model in infinite
dimensions has a transition to a commensurate AF state ($X=-1$) at the 
N\'eel temperature $T_{N}$.  $T_{N}$ increases monotonically
from zero at $U=0$ to $0.14t^*$ at $U\approx 3t^*$.  When $U$ is
increased further, $T_{N}$ decreases monotonically to zero
at $U=\infty$.  As the system is doped away from half filling,
the N\'eel temperature drops until a critical filling is reached
where the commensurate SDW becomes incommensurate.  This is shown in
Figs.~2a and 2b for the weak-coupling and strong-coupling results 
respectively.   As the system is doped further away from
half filling, the wavevector of the ordered phase changes
continuously with the electron concentration until $T_c$ drops
to zero at the incommensurate-paramagnetic phase boundary.  
The shape of the magnetic
phase boundary changes continuously from a BCS-like curve
(as a function of doping) at weak coupling to an almost linear 
curve at strong coupling, with the same crossover region $(U\approx 3t^*)$
as found for $T_{N}$ at half filling.

Phase diagrams in the weak-coupling regime ($U/t^*\le 3)$ have been
obtained from both QMC simulations and from the
theory of Eq.~(\ref{eq: mod_stoner}).  They are plotted in Fig.~2(a) for four
values of $U/t^*$ ($U/t^*=1,1.5,2,3$).   Let $X_{max}$ denote the largest
value of the scalar parameter $X$ with which incommensurate order is found
for each value of $U$.  In the QMC simulations, we find that both $X_{max}$
increases, and that the ratio of the transition temperature
at the commensurate-incommensurate phase boundary $T_I$ to the
N\'eel temperature at half filling $T_{N}$ decreases, as the
coupling strength increases. 
However, when the modified Stoner criterion is used, we find that the transition
temperature curves scale with the coupling strength and maintain the 
same approximate  shape.  This implies that $X_{max}$ will increase, while
the ratio $T_I/T_{N}$ remains constant, $T_I/T_{N}\approx 0.57$.

QMC results in the strong-coupling regime $(U/t^*\ge 3)$, along with Li 
and d'Ambrumenil's approximation 
for $T_c$ are plotted in Fig.~2(b) for $U/t^*=3,4,5,7$.  The approximate
results are generated with the restriction that only ordered states with 
$X<0$ are considered (which suppresses the ferromagnetism due to
phase separation). Under this assumption, 
the approximate strong-coupling theory predicts no incommensurate order 
(near $X=-1$), and the transition temperature curves also maintain the 
same shape as the coupling strength changes.  Accurate simulations at very 
large values of $U$ are not possible with the QMC. Thus, we are unable 
to determine whether $X_{max}$ continues to increase
in the strong-coupling regime, nor are we able to determine
what happens to $T_I/T_N$.  Of course, the strong-coupling approximation
predicts both $X_{max}$ and $T_I/T_N$ are constants, since it does not display 
any incommensurate order.

Comparison of both weak-coupling and strong-coupling approximations
with the QMC solutions 
shows that the approximate methods are unable to reproduce the qualitative
change in shape of the finite-temperature phase diagrams as a function of $U$. 

The ``phase diagram'' which indicates the commensurate-incommensurate
phase boundary (occurring at $T=T_I$) and the
incommensurate-paramagnetic phase boundary (occurring at $T=0$) is
presented in Fig.~3.  The thin (thick) solid lines denote the
commensurate-incommensurate boundary for the Stoner (modified Stoner)
criterion; the thin (thick) dashed lines
plot the corresponding paramagnetic phase boundary.  The dotted
line is the commensurate-paramagnetic phase boundary in the strong-coupling
theory, and the dots are the QMC results.  The quantum
fluctuations strongly renormalize the HF phase boundary to produce 
good agreement with the QMC.  The value of $X_{max}=0$
occurs at $U=\infty$ when the modified Stoner criterion is used, whereas
$X_{max}$ increases to the ferromagnetic point $(X=1)$ in the Stoner
theory (ferromagnetism does occur at large values of $U$ and small
electron concentration
when the modified Stoner criterion is used).  Note that the
changes to the phase diagram brought about by the fluctuations are simple: 
the critical value of $U$ is shifted by $1/U=1/U_{HF}-\chi^0(0,T_c)$, 
which pushes some of the anomalous behavior of the HF approximation 
beyond $U=\infty$. The QMC calculations show that many-body effects actually 
favor the formation of incommensurate phases, since the region of stability
of incommensurate magnetism is larger than what the weak-coupling
or strong-coupling theories would predict.

\paragraph*{Comparison to Experiment.}
Finally, it is of interest to compare our results with what is known about
magnetism in elemental Cr.  Electronic band structure
calculations~\cite{fawcett} show that the $d$-electron concentration
for Cr is 4.6/atom which is close to a half-filled band.
Doping with Mn adds an electron to the 
$d$-bands, and doping with V removes an electron.  
The commensurate-incommensurate phase boundary lies at
a doping of 0.3\% Mn and the paramagnetic phase boundary at
a doping of 3.5\% V \cite{fawcett,mackintosh}.  Since the density of states for
Cr is peaked near the band edges, rather than the band center, it is difficult
to map directly onto the Gaussian density of states (of the single-band
model in infinite dimensions).
Instead, we compare the ratio of the incommensurate ordering wavevector
to the commensurate wavevector in order to estimate the magnitude of the
Coulomb interaction for the effective single-band model.  The smallest value
for the ratio of the incommensurate wavevector to the commensurate wavevector
is 0.92 for Cr \cite{fawcett,mackintosh} implying $X_{max}=\cos (0.92\pi )=
-0.97$ for the
single-band model.  The approximate value of $U$ is then estimated to be
$U\approx 1.9t^*$ ($U\approx 1.5t^*$) for the weak-coupling theory (QMC).
(We also found that the incommensurate wavevector changes very rapidly with
doping near the commensurate-incommensurate phase boundary which is
reminiscent of the first-order jump in the wavevector that is seen in Cr.)
The ratio of $T_I$ ($325K$) to $T_{N}$ ($700K$) is 0.46
which is smaller than the weak-coupling value of 0.57 and is 
consistent with the assignment of a small value to $U/t^*$.
The N\'eel temperature at half filling is approximately 0.096 $t^*$
(0.086 $t^*$) in the weak-coupling theory (QMC)
which yields an effective bandwidth $W\equiv 4t^*=2.4$~eV 
($2.7$~eV) for the
single-band model.  This is a reasonable number since Cr has 
a bandwidth of 6.8~eV and the pileup of the density of
states at the band edges~\cite{fawcett}, 
implies that the bandwidth for the
effective single-band model must be larger than the naive
approximation of one-fifth of the total bandwidth.  

\paragraph*{Conclusion.}
We have shown that incommensurate SDW order exists
in the infinite-dimensional Hubbard model as one dopes away from
half filling.  A simple modification (due to quantum
fluctuations) of the usual Stoner criterion produces good
agreement with the weak-coupling QMC results and is easy to implement in
arbitrary dimensions (the {\bf q}-dependent susceptibility is
reduced by the local susceptibility before applying the Stoner
criterion). Since these quantum fluctuations produce large renormalizations
of the magnetic phase boundaries, it is worthwhile to repeat previous
Hartree-Fock calculations \cite{penn,schulz},  and employ the modified Stoner 
criterion, to more accuratly determine the phase diagram in the weak-coupling 
regime.  We have also found that strong-coupling theories (which
map onto effective Falicov-Kimball models) display pathological behavior in
the magnetic transition temperature, because the (approximate) irreducible
vertex functions do not reproduce the atomic limit when 
$U/t^*\rightarrow\infty$.  Finally, we have shown that much of the 
behavior found in Cr and dilute Cr alloys can be described by an 
effective single-band Hubbard model that does not include any of the microscopic
details of the bandstructure.

We would like to acknowledge useful discussions with 
B.\ Goodman,
L.\ Falicov, 
V.\ Jani\v s,
Th.\ Pruschke, 
R.\ Scalettar, 
D.\ Scalapino, 
P.\ van Dongen, 
D.\ Vollhardt,
and J.\ Wilkins.  
This work was
supported by the National Science Foundation grant Nos. DMR-9107563
and PHY-8904035
and by the Office of Naval Research grant No. N00014-93-1-0495.
In addition MJ would like to acknowledge the support of the
NSF NYI program.  Computer support was provided by the Ohio
Supercomputer Center.

$^{\dagger}$ Address after Sept. 1, 1994: Department of Physics, Georgetown
University, Washington, DC, 20057-0995.

\begin{figure}[htb]
\caption{The magnetic susceptibility for all $X({\bf{ q}})$
at various temperatures when $U=4$ and $\rho_e=0.825$.
The susceptibility displays a peak
at $X\approx -0.90$.  As shown in the inset, the transition temperature 
($T_c=0.0148 t^*$) was inferred from extrapolation of the peak inverse 
susceptibility.}
\end{figure}

\pagebreak

\begin{figure}[htb]
\caption{Phase diagram of the Hubbard model in the (a)
weak-coupling
regime ($U/t^*=1,1.5,2,3$) and (b) strong-coupling regime
($U/t^*=3,4,5,7$).  The solid (open) dots denote
the transition temperature to a commensurate (incommensurate)
SDW phase as determined by a QMC calculation.  The solid (dotted)
lines denote the transition temperature to a commensurate
(incommensurate) SDW phase using the modified Stoner criterion in
Figure 2~(a) and using Li and d'Ambrumenil's approximation in
Figure 2~(b).  The dashed lines are a fit of the QMC data to the form 
$T_c/t^*= a(x-b)^c$.  The exponent $c$ increases with increasing $U$.}
\label{fig:1}
\end{figure}

\pagebreak

\begin{figure}
\caption{``Phase diagram'' for the Hubbard model as a function
of electron concentration and $U$.  The thin (thick) solid
lines denote the commensurate-incommensurate phase boundary
for the Stoner criterion (modified Stoner criterion); the thin
(thick) dashed lines are the corresponding results for the 
incommensurate-paramagnetic phase boundary.  The dotted line is
the strong-coupling approximation for the commensurate-paramagnetic
phase boundary.  The solid (open) dots denote the QMC solutions
that display commensurate (incommensurate) SDW order.}
\label{fig:2}
\end{figure}

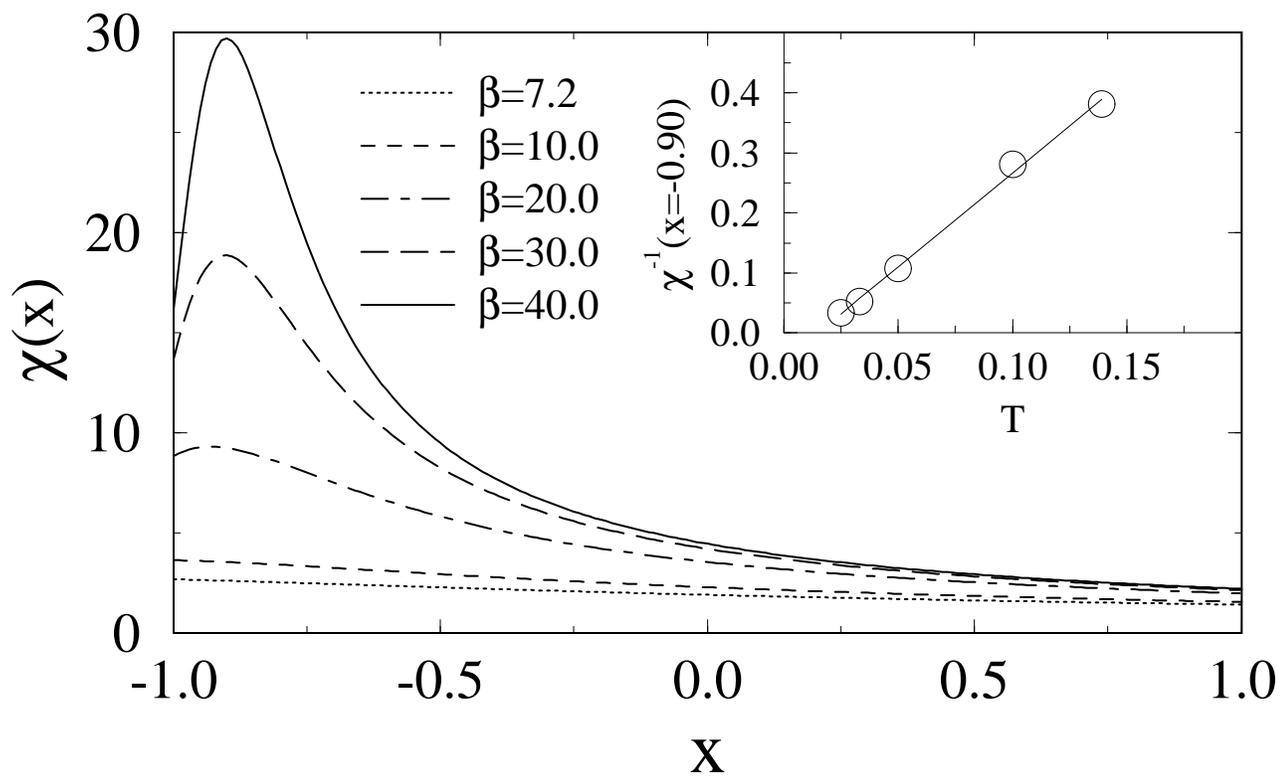
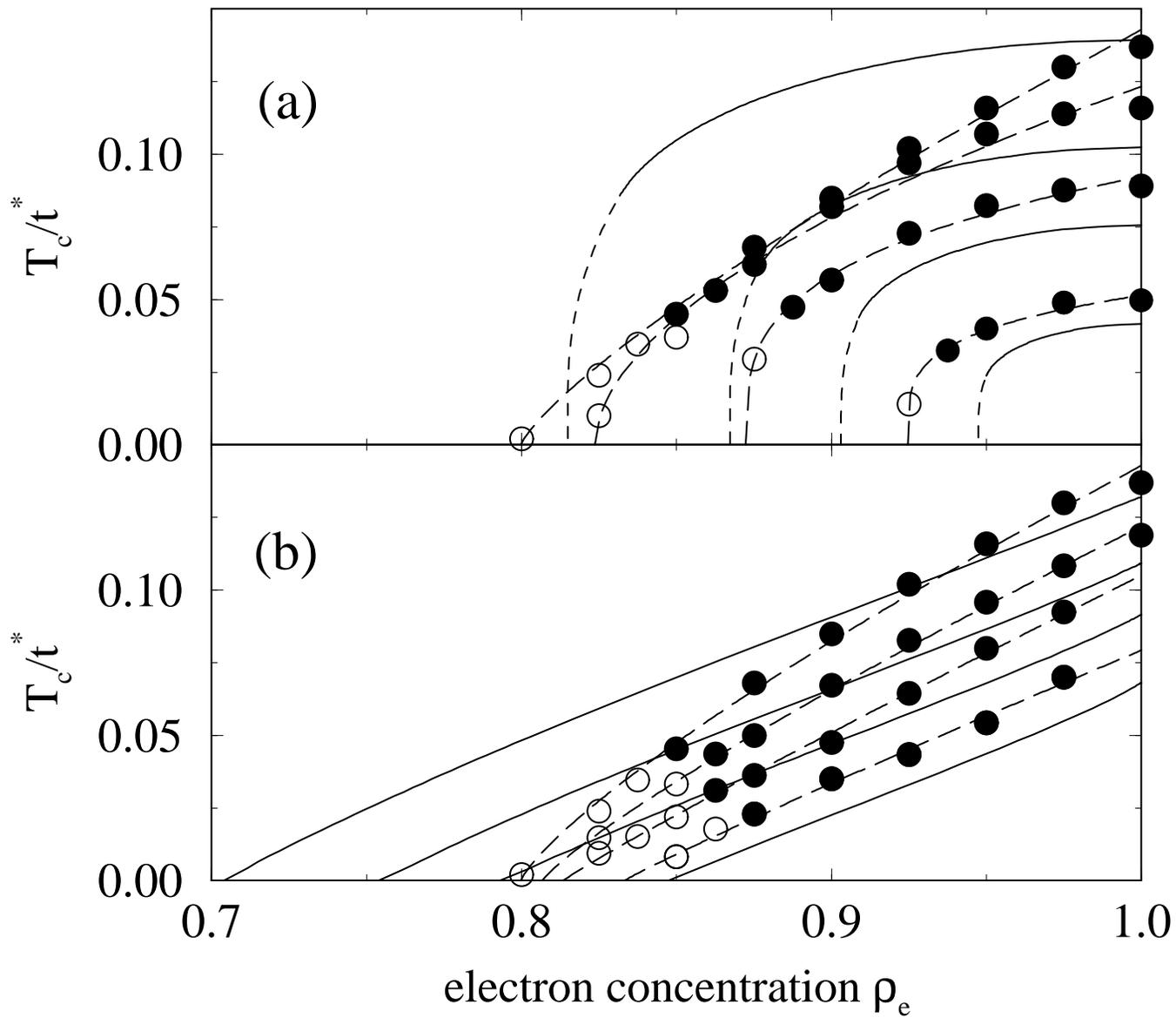
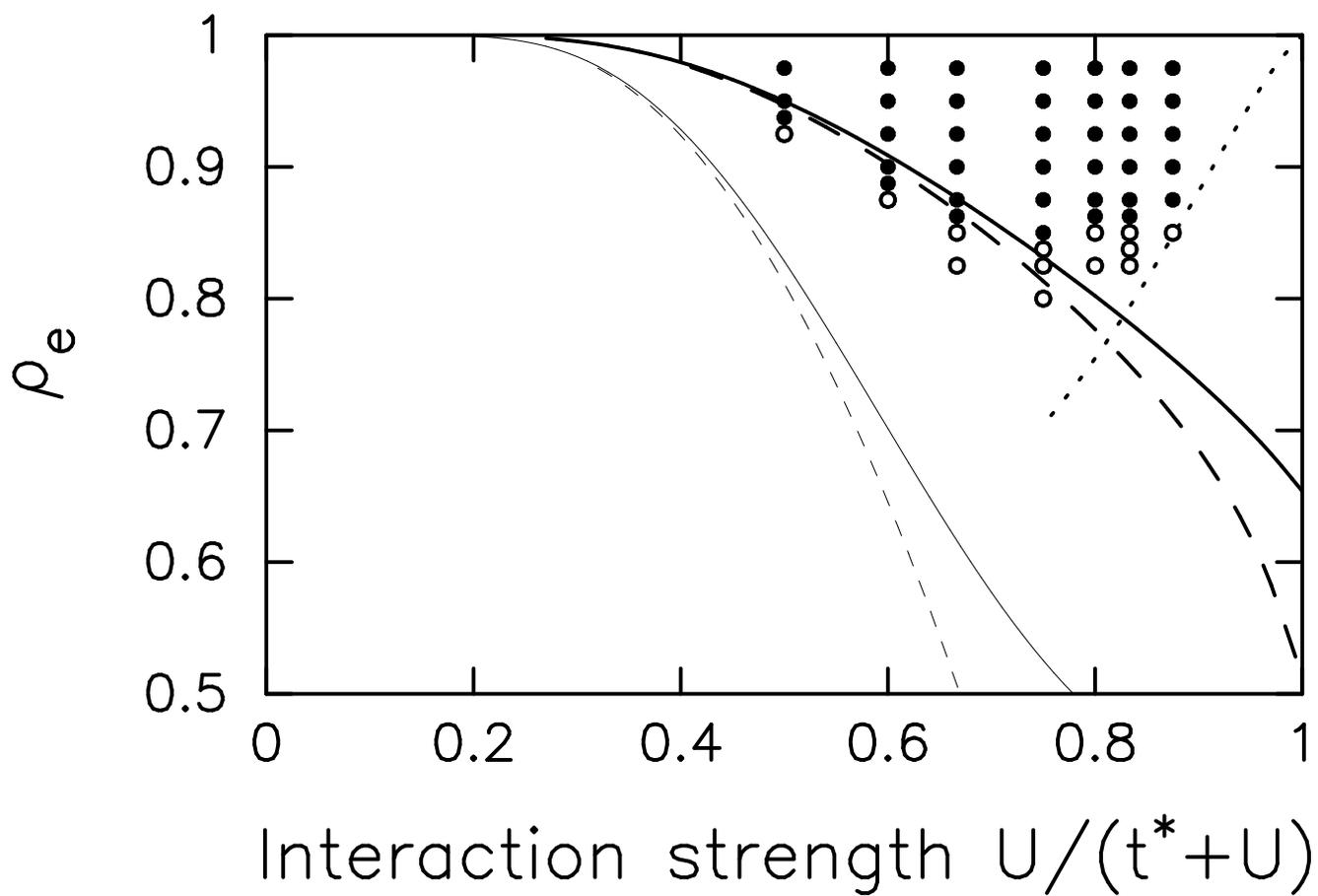
\end{document}